# A Reliable and Efficient Detection Pipeline for Rodent Ultrasonic Vocalizations

**S. S. Anis**[1-4], D. M. Kellis[5], K. F. Kaigler[5,6], M. A. Wilson[5,6], and **C. O'Reilly**[1-4]

[1] Artificial Intelligence Institute, University of South Carolina, Columbia, SC, USA
[2] Institute for Mind and Brain, University of South Carolina, Columbia, SC, USA
[3] Carolina Autism and Neurodevelopment Research Center, University of South Carolina, Columbia, SC, USA
[4] Department of Computer Science and Engineering, University of South Carolina, Columbia, SC, USA
[5] Department of Pharmacology, Physiology & Neuroscience, University of South Carolina School of Medicine, Columbia, SC, USA
[6] Columbia VA Health Care System, Columbia, SC, USA
E-mail: sanis@email.sc.edu; christian.oreilly@sc.edu

**Summary:** Analyzing ultrasonic vocalizations (USVs) is crucial for understanding rodents' affective states and social behaviors, but the manual analysis is time-consuming and prone to errors. Automated USV detection systems have been developed to address these challenges. Yet, these systems often rely on machine learning and fail to generalize effectively to new datasets. To tackle these shortcomings, we introduce ContourUSV, an efficient automated system for detecting USVs from audio recordings. Our pipeline includes spectrogram generation, cleaning, pre-processing, contour detection, post-processing, and evaluation against manual annotations. To ensure robustness and reliability, we compared ContourUSV with three state-of-the-art systems using an existing open-access USV dataset (USVSEG) and a second dataset we are releasing publicly along with this paper. On average, across the two datasets, ContourUSV outperformed the other three systems with a 1.51× improvement in precision, 1.17× in recall, 1.80× in F1 score, and 1.49× in specificity while achieving an average speedup of 117.07×.

## 1. Introduction

Rodents, such as rats and mice, use ultrasonic vocalizations (USVs) as a form of communication in various behavioral contexts. These vocalizations, which occur at frequencies beyond the range of human hearing, have become an important tool for studying the emotional states and social behaviors of rodents [1]. For example, USVs are often emitted in response to stress, mating, or social interactions, providing valuable insights into the neural mechanisms underlying these behaviors. Understanding and analyzing USVs can thus shed light on how rodents communicate, how their behavior changes in response to different stimuli, and how these processes may relate to human neuropsychiatric disorders such as anxiety, depression, post-traumatic stress disorder (PTSD), and autism [1].

USVs are high-frequency sounds (20-120 kHz) emitted by rodents in various behavioral contexts (e.g., mating, aggression, and distress) [1]. Traditional manual analysis of USVs is time-consuming, labor-intensive, and error-prone. Additionally, manual methods suffer from subjectivity, with different researchers potentially annotating the same data differently [2]. Automated systems can address these limitations by accelerating the analysis of large datasets, improving consistency, and minimizing the need for manual annotation. Several systems have been developed using signal processing and machine learning techniques [3-9]. However, these systems often fail to generalize to new datasets and sometimes require extensive manual intervention [10]. Moreover, training deep learning models can be time-consuming and resource-intensive, leaving a significant carbon footprint.

We propose a novel automated and energy-efficient approach for detecting rodent USVs using robust contour detection on spectrograms. We evaluated the reliability of our system on two open-access datasets and compared its performance to state-of-the-art methods (DeepSqueak [4], Joseph the mouse (JTM) [6], and USVSEG [9]). Through rigorous experimental analysis, the ContourUSV detection pipeline demonstrated robustness across different datasets, offering a reliable and scalable solution for large-scale USV analysis.

## 2. Related Works

Several automated systems have been developed to detect USVs. MUPET [3], an open-source software, uses signal processing techniques for rapid, unsupervised analysis of mouse USVs. It provides automated discovery and comparison of syllable types (i.e., call types) across strains and social conditions. It also incorporates noise removal and time-stamping features to facilitate behavioral analysis. DeepSqueak [4] automates USV detection and classification using deep neural networks, clustering, and supervised classification. JTM [6] proposes two alternative techniques to detect USVs: Morphological Geodesic Active Contour (GAC) [11] and Faster R-CNN [12]. While GAC offers a configurable, non-trainable approach, Faster R-CNN uses neural networks to learn from annotated data. HybridMouse [7] uses a combination of convolutional and recurrent neural networks for automatic USV identification and annotation, outperforming DeepSqueak in recall and F1 score metrics. VocalMat [5] provides tools for USV detection and classification, emphasizing user customization. It supports supervised and unsupervised methods but requires significant manual intervention. A-MUD [8] is an algorithm designed to detect mouse USVs automatically using the STx acoustic software. A-MUD achieves lower error rates than commercial software and accelerates USV detection 4 to 12 times compared to manual annotations. USVSEG [9] uses signal processing techniques to detect USV segments amid noise and track spectral peaks for syllables. The performance of this system has been demonstrated across several rodent species on an open-access dataset of the same name. BootSnap [10] can classify calls in syllables but does not propose a USV detection method, relying instead on existing detectors. The authors of BootSnap compared multiple detection methods, including DeepSqueak, USVSEG, A-MUD, and MUPET. They found A-MUD and USVSEG to have higher true positive rates and A-MUD to have lower false detection rates. A study of multiple deep-learning algorithms for neonatal murine USV detection demonstrated high performance on an open-access dataset of recordings [13]. However, this dataset does not include manual annotations. Therefore, we could not use it for comparative analysis. Despite these advances, the accuracy and reliability of these systems across datasets and experiments remain understudied. Further, most of these systems use advanced machine learning approaches for a task that might be solvable by lighter, more predictable signal processing approaches. Occam's Razor suggests it might be advantageous to look at computationally simpler and lighter approaches if they perform at least similarly.

## 3. Datasets

To assess the performance of our approach, we released an open-access dataset called USCMed and utilized another open-access dataset, USVSEG.

### 3.1. USCMed Dataset

The USCMed Dataset was collected at the University of South Carolina School of Medicine. This dataset involves a study designed to examine individual differences in rat fear conditioning and extinction. During this protocol, we recorded the USVs from male Long Evans rats (N=27) exposed to the following four experimental conditions (Figure 1; [14], [15]):

1) <u>Fear Acquisition:</u> Three light foot shocks paired with 10-second 2 kHz tones and separated by a one-minute inter-stimulus interval were administered.
2) <u>Contextual Fear:</u> Recording in the same environment (cage), but without tones or shocks.
3) <u>Cued Fear Extinction:</u> Twenty tones presented at one-minute intervals without co-occurring shocks, in a different environment.
4) <u>Extinction Recall:</u> Same as for 3), but two days later.

The audio was recorded at 250 kHz with UltraVox XT (Noldus Information Technology, Inc., Leesburg, VA). Each recording was manually annotated for the USV call start and stop times, frequency at max amplitude, and mean amplitude. These manual annotations served as the gold standard for evaluating our ContourUSV detector. Along with this paper, we released a subset of the USCMed dataset with the audio recordings (N=27) and gold standard annotations for the Context trial. The USCMed dataset is openly available at https://doi.org/10.5281/zenodo.14211069.

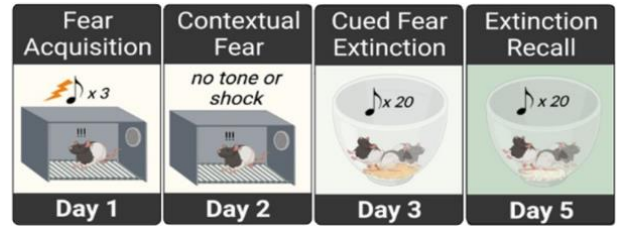

**Fig. 1.** Paradigm used to collect the USCMed dataset (created with BioRender.com).

### 3.2. USVSEG Dataset

The USVSEG [9] dataset consists of recordings from mice (N=20; C57BL/6J, BALB/c, and Shank2- adult males, and juvenile C57BL/6J mice), rats (N=7; adult females, in distressing and pleasant contexts), and gerbils (N=2), recorded at 250 kHz using a commercial condenser microphone and an A/D converter (UltraSoundGate, Avisoft Bioacoustics, Berlin, Germany; SpectoLibellus2D, Katou Acoustics Consultant Office, Kanagawa, Japan). The USVSEG dataset includes manual USV annotations and is openly available at https://doi.org/10.5281/zenodo.3428023.

## 4. ContourUSV Detection Pipeline

This section describes the architecture of the ContourUSV pipeline as shown in Figure 2. The code for ContourUSV is available on GitHub (https://github.com/lina-usc/contourusv).

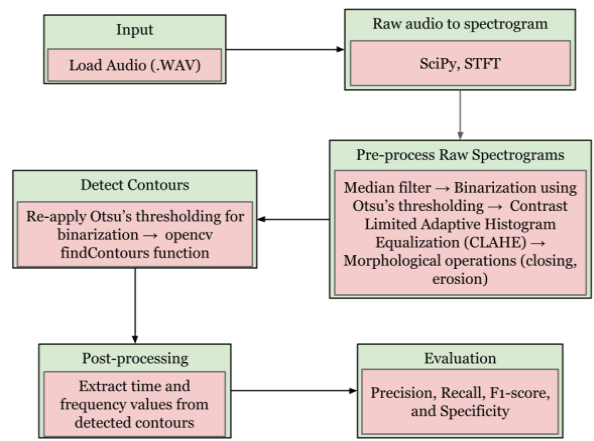

**Fig. 2.** Architecture of the ContourUSV detection pipeline.

### 4.1. Spectrogram Generation

The initial step in our pipeline involves generating spectrograms from raw audio recordings. These spectrograms serve as the basis for the subsequent processing and detection. We first read the audio signals from the .wav files (Figure 3) using the Python SciPy library [16] and transform these signals into spectrograms (i.e., time-frequency representations) using the Short-Time Fourier Transform (STFT) implemented in MNE-Python [17]. The STFT was computed using a window size of 2,500 samples and a time step of 5 ms. To ensure the pipeline is not sensitive to differences in sample rate, we used the resample function from MNE-Python to resample signals to 250 kHz, if necessary. This function uses the same approach as SciPy, relying on the Fast Fourier Transform. We focused on the 15-115 kHz frequency range to isolate relevant signal components. The frequency resolution (100 Hz) is determined by the window size and the sampling frequency. The result is a 2D NumPy array representing the spectrogram data (time-frequency representation). Then, the NumPy array is passed to the spectrogram pre-processing and cleaning stage.

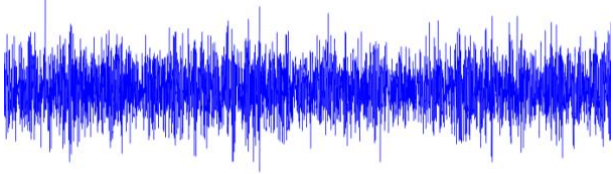

**Fig. 3.** Example audio signal loaded from a .wav file (8s).

For visualization, we tested various Matplotlib [18] colormaps, and chose 'viridis' as it provided the best visual clarity for analyzing USVs (Figure 4).

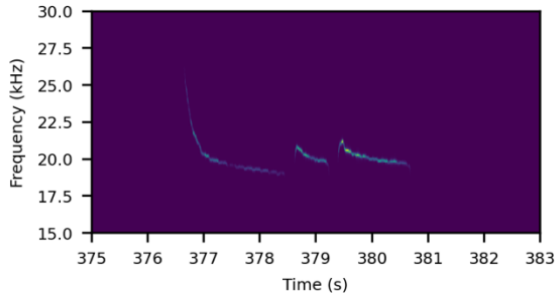

**Fig. 4.** Raw spectrogram from audio recording (8s).

### 4.2. Pre-processing and Cleaning

The raw spectrogram data is further pre-processed to enhance the contrast between the USVs and the background noise. First, we apply a median filter from SciPy to the spectrogram data. Median filtering is used to reduce noise in the image while preserving edges. It replaces each pixel's value with the median value in its neighborhood. Then, we normalize the spectrogram using OpenCV [19] to scale the pixel intensity values of the filtered image to the [0, 255] range. The normalization function then casts the result to an 8-bit unsigned integer type, as standard for grayscale (single channel) images and required for OpenCV thresholding functions [18]. Next, we apply a thresholding approach to binarize the spectrogram. We chose Otsu's thresholding [20] instead of global thresholding to dynamically determine the optimal threshold for consistent binarization of spectrograms. This eliminates the need for a fixed threshold, enhancing robustness against noise and ensuring a clear separation of USVs from the background for reliable contour detection. To improve the visibility of the USVs, a contrast-limited adaptive histogram equalization (CLAHE) [21] is applied to the binary spectrograms. This technique enhances the local contrast of the images and makes the USV contours more prominent by applying an adaptive histogram equalization with a user-defined contrast limit to prevent over-amplification of noise. Finally, we apply morphological operations [18], specifically closing, which is a combination of dilation followed by erosion, which smooths object boundaries, fills small holes, and connects close objects in the spectrograms. This step ensures that the detected USV contours are continuous and well-defined as shown in Figure 5.

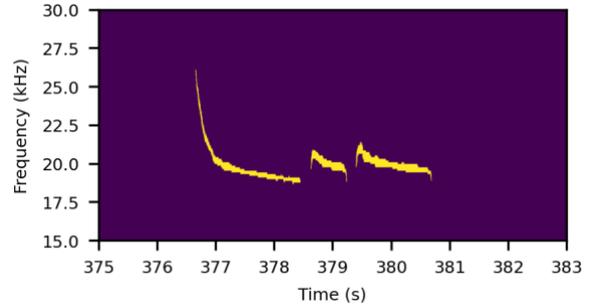

**Fig. 5.** Cleaned and pre-processed spectrogram.

### 4.3. Contour Detection

The contour detection stage focuses on identifying and extracting the contours of USVs from the pre-processed spectrogram image data. This critical step allows for the precise localization of USVs in the time and frequency domains. In the pre-processing steps, applying CLAHE after the first binarization results in non-binary spectrograms. Otsu's thresholding is applied again to binarize the spectrograms, providing a clear separation between the USVs and the background and allowing for more reliable contour detection.

Contours were then extracted from the thresholded binary spectrograms using the OpenCV findContours function. This function retrieves the external contours, representing the boundaries of the detected USVs. Figure 6 illustrates an example of detected USVs.

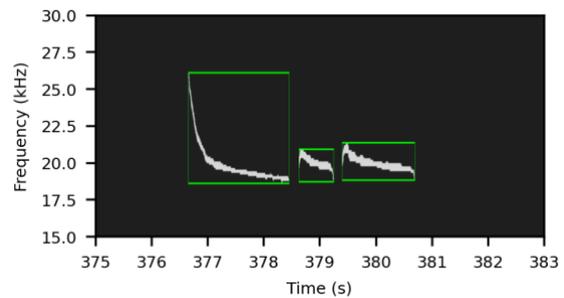

**Fig. 6.** Detected contours and bounding boxes (green).

### 4.4. Post-processing Annotations

For each detected contour, a bounding box is computed using the OpenCV boundingRect function (see Figure 6) and the time and frequency boundaries are calculated based on the position and dimensions of the bounding box relative to the image.

### 4.5. Evaluation

To assess the performance of ContourUSV, we compared the detected USVs against our gold standard (manual annotations) using the performance metrics shown in Figure 7. First, we obtained both the predicted calls (system output) and actual calls (gold standard, obtained through manual annotations). Each annotation (manual or automated) included the start and end times of detected USVs. These annotations were used to create two sets of binary labels (predicted and actual) encoding the presence or absence of USVs for every time point in the audio signal. For this process, time windows specified in the annotations were mapped to sample indices of the audio files. To evaluate the system's performance, we calculated key metrics by comparing the predicted binary labels to the gold standard (actual) labels across all time points. True positives (TP), false positives (FP), false negatives (FN), and true negatives (TN) were defined as usual (Figure 7, left panel). Using these values, we computed the precision, recall, F1 score, and specificity (Figure 7, right panel). Finally, we aggregated the results across all recordings to compute the mean and standard deviation for each metric, providing an overall performance evaluation of the detection systems across various conditions and datasets. The statistical significance of performance differences between detectors was assessed with paired t-tests.

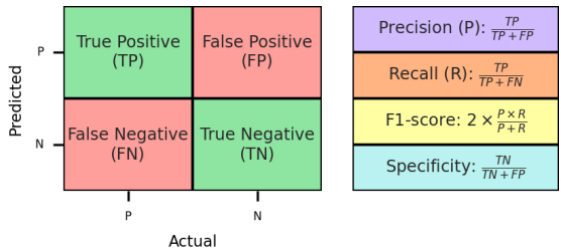

**Fig. 7.** Definition of the metrics used for assessing the performance of the different systems.

## 5. Results

This section presents the detection results for the USCMed and USVSEG datasets tested with ContourUSV, DeepSqueak, JTM, and USVSEG.

### 5.1. USCMed Dataset Results

Tables 1 and 2 present the results on the USCMed Dataset for the Fear Acquisition and the Context conditions, respectively. Bold is used to indicate statistical significance when comparing the performances of the best vs. second-best detectors. Table 3 and Table 4 show the paired t-test results comparing the best-performing model with the other models for Fear Acquisition and Context conditions, respectively. Statistical significance is defined at p < 0.05.

**Table. 1.** USCMed dataset fear acquisition trial results (Mean ± SD).

| Metric | DeepSqueak | JTM | ContourUSV | USVSEG |
|---|---|---|---|---|
| Precision | 0.23 ± 0.15 | 0.77 ± 0.35 | **0.86 ± 0.23** | 0.42 ± 0.24 |
| Recall | **0.97 ± 0.05** | 0.59 ± 0.17 | 0.76 ± 0.23 | 0.89 ± 0.10 |
| F1 Score | 0.35 ± 0.19 | 0.62 ± 0.23 | **0.80 ± 0.22** | 0.53 ± 0.26 |
| Specificity | 0.22 ± 0.14 | 0.94 ± 0.16 | **0.99 ± 0.01** | 0.71 ± 0.16 |

**Table. 2.** USCMed dataset context trial results (Mean ± SD).

| Metric | DeepSqueak | JTM | ContourUSV | USVSEG |
|---|---|---|---|---|
| Precision | 0.17 ± 0.11 | 0.99 ± 0.01 | **0.99 ± 0.01** | 0.51 ± 0.17 |
| Recall | **1.00 ± 0.00** | 0.22 ± 0.09 | 0.87 ± 0.05 | 0.42 ± 0.10 |
| F1 Score | 0.28 ± 0.15 | 0.36 ± 0.12 | **0.93 ± 0.02** | 0.44 ± 0.08 |
| Specificity | 0.21 ± 0.13 | 1.00 ± 0.00 | **1.00 ± 0.00** | 0.95 ± 0.01 |

**Table. 3.** Fear acquisition trial paired t-test results comparing the best-performing model with other models for each metric.

| Metric (Best Model) | Comparison | T-value | P-value |
|---|---|---|---|
| Precision (ContourUSV) | DeepSqueak | 17.80 | **1.04e$^{-15}$** |
| | JTM | 2.09 | **4.69e$^{-2}$** |
| | USVSEG | 11.64 | **1.37e$^{-11}$** |
| Recall (DeepSqueak) | ContourUSV | 4.51 | **1.32e$^{-4}$** |
| | JTM | 11.34 | **2.37e$^{-11}$** |
| | USVSEG | 4.28 | **2.42e$^{-4}$** |
| F1 Score (ContourUSV) | DeepSqueak | 12.98 | **1.31e$^{-12}$** |
| | JTM | 4.90 | **4.86e$^{-5}$** |
| | USVSEG | 7.37 | **1.02e$^{-7}$** |
| Specificity (ContourUSV) | DeepSqueak | 28.39 | **1.53e$^{-20}$** |
| | JTM | 1.66 | 1.09e$^{-1}$ |
| | USVSEG | 9.23 | **1.57e$^{-9}$** |

**Table. 4.** Context trial paired t-test results comparing the best-performing model with other models for each metric.

| Metric (Best Model) | Comparison | T-value | P-value |
|---|---|---|---|
| Precision (ContourUSV) | DeepSqueak | 30.73 | **2.63e-18** |
| | JTM | 3.60 | **1.77e-03** |
| | USVSEG | 11.98 | **1.39e-10** |
| Recall (DeepSqueak) | ContourUSV | -10.83 | **8.09e-10** |
| | JTM | 38.80 | **2.66e-20** |
| | USVSEG | 25.41 | **1.07e-16** |
| F1 Score (ContourUSV) | DeepSqueak | 22.53 | **1.10e-15** |
| | JTM | 25.10 | **1.36e-16** |
| | USVSEG | 32.26 | **1.01e-18** |
| Specificity (ContourUSV) | DeepSqueak | 26.08 | **6.46e-17** |
| | JTM | -4.15 | **4.98e-04** |
| | USVSEG | 17.83 | **9.53e-14** |

The computation time was also recorded for each model on the USCMed dataset. ContourUSV performed detection in 359 seconds (i.e., 6 minutes) which is 7.26× faster than DeepSqueak's execution time of 2,607 seconds (i.e., more than 43 minutes), 226.89× faster than JTM's execution time of 81,462 seconds (i.e., more than 22 hours), and 51.32× faster than USVSEG's execution time of 18,426 seconds (i.e., more than 5 hours).

### 5.2. USVSEG Dataset Results

Table 5 presents the evaluation metrics for different species on the USVSEG dataset. The paired t-test results for these comparisons are presented in Table 6 for all species.

Since ControurUSV has been designed and tested in the context of a fear acquisition protocol, we focused on detecting rat distress calls. To test if the lower performance of this system on the USVSEG dataset could be due to a failure to detect 50 kHz (positively valenced) calls, we additionally tested the detectors on only the 22 kHz calls from the USVSEG dataset. In this context, ContourUSV achieved an F1 Score of 0.96 ± 0.03, while USVSEG, DeepSqueak, and JTM obtained

F1-scores of 0.97 ± 0.01, 0.40 ± 0.22, and 0.55 ± 0.17, respectively.

Table. 5. ContourUSV evaluation metrics on the USVSEG dataset.

| Metric | DeepSqueak | JTM | ContourUSV | USVSEG |
|---|---|---|---|---|
| **Gerbil** | | | | |
| Precision | 0.17 ± 0.00 | **0.97 ± 0.02** | 0.66 ± 0.17 | 0.72 ± 0.04 |
| Recall | **0.99 ± 0.05** | 0.28 ± 0.21 | 0.97 ± 0.01 | 0.97 ± 0.03 |
| F1 Score | 0.29 ± 0.01 | 0.42 ± 0.26 | 0.78 ± 0.12 | 0.82 ± 0.04 |
| Specificity | 0.04 ± 0.02 | **1.00 ± 0.00** | 0.89 ± 0.08 | 0.92 ± 0.01 |
| **Mouse** | | | | |
| Precision | 0.10 ± 0.10 | 0.62 ± 0.45 | 0.39 ± 0.27 | **0.73 ± 0.14** |
| Recall | **0.99 ± 0.02** | 0.16 ± 0.19 | 0.59 ± 0.17 | 0.91 ± 0.05 |
| F1 Score | 0.18 ± 0.14 | 0.23 ± 0.25 | 0.42 ± 0.23 | **0.80 ± 0.09** |
| Specificity | 0.09 ± 0.09 | 0.69 ± 0.47 | 0.89 ± 0.08 | **0.97 ± 0.03** |
| **Rat** | | | | |
| Precision | 0.24 ± 0.12 | **0.98 ± 0.02** | 0.84 ± 0.29 | 0.95 ± 0.03 |
| Recall | **1.00 ± 0.00** | 0.24 ± 0.23 | 0.84 ± 0.17 | 0.90 ± 0.07 |
| F1 Score | 0.37 ± 0.16 | 0.35 ± 0.28 | 0.81 ± 0.24 | 0.93 ± 0.05 |
| Specificity | 0.06 ± 0.06 | **1.00 ± 0.00** | 0.96 ± 0.07 | 0.99 ± 0.01 |
| **All Species** | | | | |
| Precision | 0.14 ± 0.11 | 0.73 ± 0.41 | 0.52 ± 0.33 | 0.78 ± 0.15 |
| Recall | **0.99 ± 0.02** | 0.19 ± 0.20 | 0.68 ± 0.21 | 0.91 ± 0.06 |
| F1 Score | 0.23 ± 0.17 | 0.27 ± 0.26 | 0.54 ± 0.28 | **0.83 ± 0.10** |
| Specificity | 0.08 ± 0.08 | 0.79 ± 0.41 | 0.91 ± 0.08 | **0.97 ± 0.03** |

Table. 6. Paired t-test results comparing the best-performing model with other models for each metric on the USVSEG dataset (all species).

| Metric (Best Model) | Comparison | T-value | P-value |
|---|---|---|---|
| Precision (USVSEG) | DeepSqueak | -28.24 | **4.04e⁻²²** |
| | JTM | -0.67 | 5.09e⁻⁰¹ |
| | ContourUSV | -5.20 | **1.62e⁻⁰⁵** |
| Recall (DeepSqueak) | ContourUSV | 8.25 | **5.67e⁻⁰⁹** |
| | JTM | 22.12 | **2.85e⁻¹⁹** |
| | USVSEG | 7.22 | **7.38e⁻⁰⁸** |
| F1 Score (USVSEG) | DeepSqueak | -26.06 | **3.54e⁻²¹** |
| | JTM | -13.79 | **5.29e⁻¹⁴** |
| | ContourUSV | -6.58 | **3.89e⁻⁰⁷** |
| Specificity (USVSEG) | DeepSqueak | -64.55 | **5.22e⁻³²** |
| | JTM | -2.29 | 2.96e⁻⁰² |
| | ContourUSV | -4.12 | **3.02e⁻⁰⁴** |

## 6. Discussion

Our primary goal was to develop an automated approach for detecting USVs in rodents that addresses the limitations of existing methods in terms of reliability, accuracy, simplicity, and computational demand. This focus on efficiency and simplicity is particularly important for applications that require real-time or embedded implementation. For instance, embedding such a detector in hardware could enable the *online* detection of USVs during an experiment, allowing researchers to dynamically modify experimental conditions based on rodent vocalizations. Such functionality could be especially beneficial in biofeedback applications, where immediate responses to vocalizations are crucial, such as in studies of social interaction or stress response. Additionally, a computationally lightweight system facilitates deployment in resource-constrained environments, such as portable devices for field studies. Since every second of recordings takes about 0.057s to process with ContourUSV on a modern laptop (MacBook Pro), the computational efficiency of this approach would permit such real-time USV detection. Although we did not design ContourUSV for such a purpose, only minimal modifications should be required to adapt the code for such applications. However, we did not test ContourUSV's performance for online detection, a topic that would require additional research.

When we initiated this project, we also wanted this detector to be open-source and not require proprietary software (e.g., Matlab). Furthermore, the limited availability of high-quality USV datasets with reliable gold standard annotations is one of the key challenges faced in our experiments as well as in this research field. Hence, we released a subset of our dataset (including the manual annotations) publicly to support USV detection benchmarking in future studies.

In our comparative analysis, ContourUSV had a high F1 score for both datasets, demonstrating strong reliability. Large performance differences between datasets highlight the significant effect that dataset properties have on the effectiveness of these systems. Since we used the USCMed Dataset for development, ContourUSV may have an unfair advantage over other systems on this dataset. The same is likely true for the evaluation of the USVSEG detector on the dataset of the same name. Nevertheless, ContourUSV shows superior reliability across datasets. Benchmarking against additional datasets would be required to corroborate this superior reliability.

Moreover, for the USCMed dataset, the results published in this paper are only for male rats on the fear acquisition and context trials. Thus, this dataset contains mostly 22 kHz call types. However, experiments with female rats and other experimental conditions within this data collection protocol, which contain a wider variety of call types (e.g., 50 kHz calls), are in progress, and results for these experiments will be published in future work. We are also investigating various denoising approaches, including single-channel decompositions. We hope such additions will allow the automatic removal of various noise sources and make ContourUSV even more reliable when processing recordings with unexpected sources of artifacts.

## 7. Conclusion

The ContourUSV detection method was designed to identify and localize USVs within spectrograms generated from audio recordings. This method employs a combination of preprocessing techniques, including median filtering, otsu's thresholding, morphological operations, and contour extraction, to enhance and detect the contours of USVs. ContourUSV utilizes the OpenCV findContours function to accurately detect and annotate the temporal and frequency boundaries of each vocalization.

Our comparative analysis shows that ContourUSV consistently achieves higher mean F1 scores in the USCMed dataset. On the other hand, for the USVSEG dataset, ContourUSV performs as the second best after the USVSEG model. Since the development of ContourUSV utilized the USCMed dataset, which includes only male rats, its performance on the USVSEG dataset, comprising various rodent species, does not match that of the USVSEG model. Nevertheless, ContourUSV still outperforms both DeepSqueak and JTM in terms of F1 scores. The gap in recall and accuracy for ContourUSV is likely due to the lack of extensive noise reduction or filtering. On average, across the two datasets, ContourUSV outperformed the other three systems with a 1.51× improvement in precision, 1.17× in recall, 1.80× in F1 score, and 1.49× in specificity, while achieving an average speedup of 117.07×.

Clustering and classifying calls for syntax analyses is an important area of future development for USV analyses. Without accurately detecting the calls first, these next stages of behavior analysis based on USVs cannot bring fruitful results. Thus, the ContourUSV detection pipeline serves an important role in advancing USV-based analysis of rodent behavior.

## Acknowledgments

This work was supported by COR's startup package at USC, merit Award # I01 BX001374 to MAW from the United States Department of Veterans Affairs Biomedical Laboratory Research and Development Service (VA BLRD), and the USC VP for Research [ASPIRE II award to MAW; SPARC award to DMK]. The authors would also like to acknowledge the contributions of Sydnie L. Mick and Alicia N. Thomas to the collection of the USCMed dataset.